# Easy computation of the Bayes Factor to fully quantify Occam's razor

D.J. Dunstan,* J. Crowne, and A.J. Drew

School of Physics and Astronomy, Queen Mary University of London, London E1 4NS, UK.

* Correspondence to **d.dunstan@qmul.ac.uk**

**Abstract:** The Bayes factor is the gold-standard figure of merit for comparing fits of models to data, for hypothesis selection and parameter estimation. However it is little used because it is computationally very intensive. Here it is shown how Bayes factors can be calculated accurately and easily, so that any least-squares or maximum-likelihood fits may be routinely followed by the calculation of Bayes factors to guide the best choice of model and hence the best estimations of parameters. Approximations to the Bayes factor, such as the Bayesian Information Criterion (BIC), are increasingly used. Occam's razor expresses a primary intuition, that parameters should not be multiplied unnecessarily, and that is quantified by the BIC. The Bayes factor quantifies two further intuitions. Models with physically-meaningful parameters are preferable to models with physically-meaningless parameters. Models that could fail to fit the data, yet which do fit, are preferable to models which span the data space and are therefore guaranteed to fit the data. The outcomes of using Bayes factors are often very different from traditional statistics tests and from the BIC. Three examples are given. In two of these examples, the easy calculation of the Bayes factor is exact. The third example illustrates the rare conditions under which it has some error and shows how to diagnose and correct the error.

**Introduction**

Least-squares (LS) fitting is used in almost every area of science, for parameter estimation (uncertainties as well as values) and hypothesis selection.[1] Yet the perennial problem faced by every user of LS fitting is how to know if the fit is as good as may be, and especially how to detect over-fitting and under-fitting. The parameter values and uncertainties returned by a fit to a model cannot be trusted if the choice of model cannot be justified. There are many quantitative criteria for goodness of fit in classical statistics. However, compliance with Occam's razor – avoiding multiplying parameters unnecessarily – is best done by Bayesian methods.[2] Although they were central to early work in probability and statistics by such as Laplace and Gauss, they were largely neglected for most of the last century, and their reintroduction initially by Jeffreys[3] in 1939 was very controversial for decades. For a non-technical discussion, see Jaynes.[4] However, the BIC (Bayesian Information Criterion) and related quantities are now largely accepted and routinely used throughout science for model selection and parameter estimation.[5,6,7] These quantities are often described as quantifying Occam's razor, because they provide quantitative criteria to optimise the number of parameters. However, they are gross approximations to the Bayes Factor (BF), because they treat all parameters alike. The BF is the ratio of the probability of one hypothesis or model to the probability of another, taking into account the *prior* probabilities of the parameter values of each model, as well as the probabilities of the data under each model.[2] The BF thereby quantifies further intuitions related to Occam's razor, based on our prior knowledge of the parameters.[8] In particular, it quantifies our intuitive preferences for models with physically-meaningful parameters, against models with physically-meaningless parameters, and for models that do not span the data space yet which fit the data, against models which span the data space and are therefore guaranteed to fit the data. This is very desirable as it often gives outcomes for model selection and parameter estimation quite different from the BIC or classical statistical methods, as we see below in three examples. However, calculating the BF is computationally very demanding. It requires multi-dimensional integrals over the parameter spaces. Even for only three or four parameters, direct analytic or numerical integrals can be impossible, let alone for tens of parameters. Computationally-intensive techniques such as Markov-chain Monte Carlo integration are needed.[8] Consequently, the easily-calculated approximations such as the BIC are widely used. Here we show how the computational difficulties of the integrals may be by-passed and that the BF can be calculated very easily from the parameter covariance matrices (see Supplementary Information §1, below, p.21), which are routinely calculated in almost all LS fitting routines. The BF calculated in this way is a mathematical approximation which is usually exact. Rarely, the method overestimates the BF, giving what could be termed a rough Bayes Factor (RBF). We show how to detect and correct cases where this inexactitude arises.

LS fitting minimises the sum of the squares of the residuals. When the residuals are normally-distributed, this is equivalent to maximizing the likelihood function, $L$, to get $L_{max}$ which is the probability of the data given the fit (see Supplementary Information, §2). The BIC quantifies the primary Occam intuition (not multiplying parameters unnecessarily) by applying to $L_{max}$ for each parameter a penalty, ½ln$n$, where $n$ is the number of data points. This treats all parameters alike. The BF does not use $L_{max}$. Instead, $L$, multiplied by the parameter probability-distribution functions (pdfs), is integrated over the whole parameter space. This integral does not appear to have a name in the literature; we call it here the Bayes Factor Integral (BFI; see Supplementary Information, §3), although we obtain it without integration. The BF between two models or hypotheses is then the ratio of their BFIs. The BFI and BF have higher values for – i.e. favour – parameters whose values have narrow pdfs around their fitted values, against parameters with broad pdfs. The former are usually physically-meaningful parameters whose plausible ranges are restricted by our prior knowledge. The latter are typically polynomial or Fourier coefficients in an arbitrary background function, or the parameters of arbitrary extra peaks, which, for want of physical meaning, can have wide ranges of plausible values. See Supplementary Information, §4. The result is that the BF often gives hypothesis preferences, and parameter estimations, very different from those proposed by other methods that treat all parameters alike, including the Bayesian BIC as well as traditional statistical tests such as the *p*-statistic or the 3σ test.

In what follows, we use three examples to illustrate these differences and demonstrate the full quantification of Occam's razor now made routinely possible. Before considering the examples, however, we demonstrate the logic of the method using a two-parameter LS fit to one of the data-sets from the first example. The full mathematics is not needed to follow the logic, so it is given in the Supplementary Information, §§1-4.

**The Method**

Figure 1 illustrates the method, the difficulty, and its solution, using a two-parameter fit to a curve-fitting problem. The data, for the band-gap of GaAs as a function of pressure, $E_g(P)$, from Goñi *et al.*[10,] is shown in Fig.1a together with two LS fits; in Example 1 below we find which fit is preferable. Here we are concerned with calculating the BFI for one of them, $E_g = E_{g0} + bP + cP^2$. After doing the LS fit, the calculation starts with the integral of $L$ over the parameter space. From the LS fit, we have the fitted values $b_f$ and $c_f$ at which $L = L_{max}$. We write $L$ as a function of the residuals with $b$ and $c$ as variables, and it is easy to calculate a few values of $L$ with either $b$ or $c$ varying and the other at its fitted value. These values are plotted in Fig.1b, normalised to unity peak height, and with the abscissa axes in units of the uncertainties $\sigma_b$ and $\sigma_c$ of the parameters.

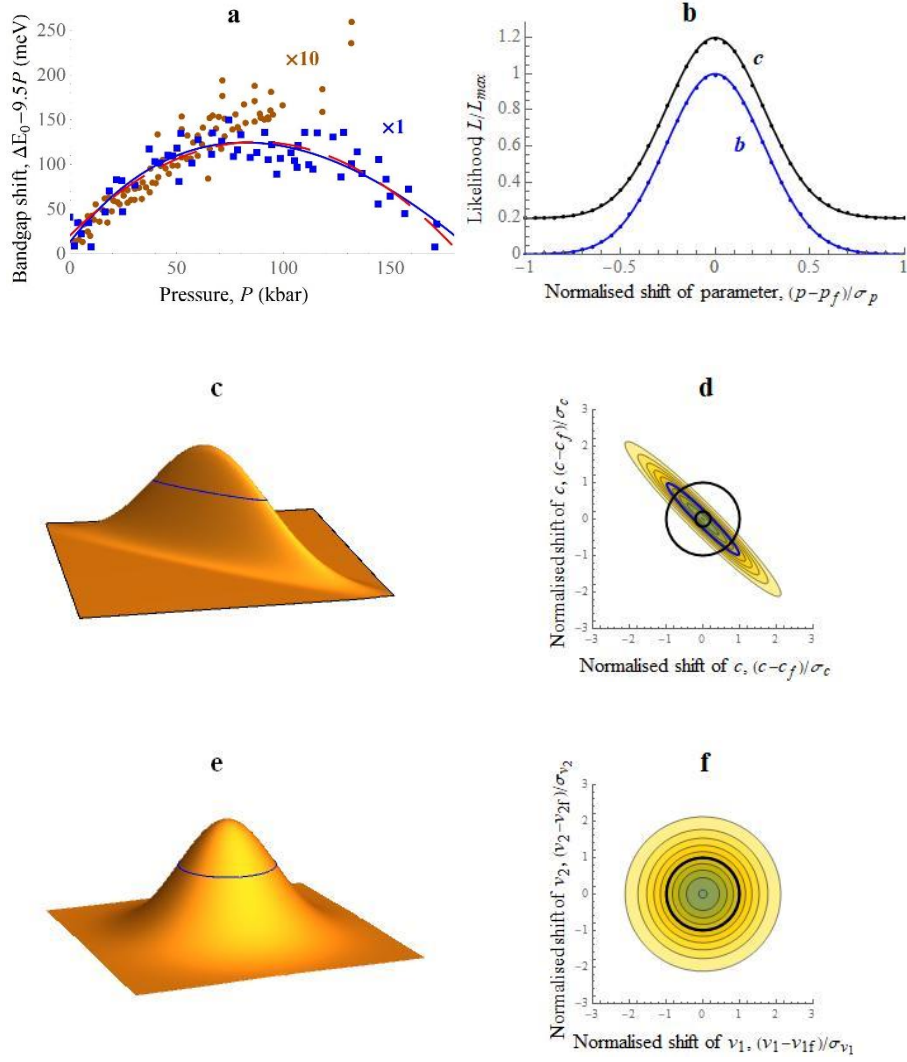

**Fig.1.** In (a), data for $E_g(P)$ in GaAs from Goñi *et al.*[10] (■) and from Perlin *et al.*[14] (●) are shown after subtraction of the straight line $E_0 + 9.5P$ to make the curvature more visible. The Perlin data is expanded ×10 on both axes for clarity. Two least-squares fits to the Goñi data are shown, polynomial (red) and Murnaghan (blue). In (b), the likelihood is plotted as a function of the polynomial parameters $p_i$ ($i = b, c$) and fitted with Gaussians of width $\delta b$ and $\delta c$. For clarity the data and fit for $c$ are shifted up by 0.2. In (c) and (d), $L(b, c)$ is plotted with normalised axes, and the heavy blue ellipse is the contour at the height $e^{-½}$. On the contour plot in (d), the area of the large circle of radius 1 corresponds to the product of the uncertainties $\sigma_b \sigma_c$. Rotating the parameter basis set to $v_1$ and $v_2$

gives the corresponding plots of (e) and (f) from which the normalised products $\sigma_{v_1}\sigma_{v_2}$ and $\delta_{v_1}\delta_{v_2}$ are equal and both correctly give the area of the contour line.

It has long been recognised that in the simple case of a one-parameter LS fit ($i = 1$), the function $L(p_1)$ is normally a Gaussian centred on the fitted value $p_{1f}$ and of height $L_{max}$.[2,9] Moreover, the half-width of this Gaussian, $\delta p_1$, is equal to the uncertainty or error returned by the LS routine for this parameter, $\sigma_1$. Consequently, the integral of $L$ over the one-dimensional parameter space is just the area of the Gaussian. This is normally exact – there is no need to carry out the explicit integral.[2,9] However, for two parameters we cannot simply multiply the two uncertainties. Notice that the widths of the two Gaussians of Fig.1b, $\delta b$ and $\delta c$, are much less than the uncertainties $\sigma_b$ and $\sigma_c$ by which the *x*-axis is normalised. Clearly using the product $\sigma_b\sigma_c$ will give a gross overestimate of the two-dimensional integral of $L$.

Plotting the full function $L(b, c)$ (Fig.1c) shows why this happens. The function is Gaussian about its peak in all directions, but it is skew to the axes. The skewness expresses the parameter covariance. Here, if the random errors in the data were to put, e.g., the initial gradient *b* higher, *c* will be fitted with a higher value as well. This corresponds here to positive off-diagonal elements in the correlation and covariance matrices. To obtain the volume of $L(b, c)$, we need the area of the elliptical contour shown in Fig.1c. The contour map of $L(b, c)$ (Fig.1d) shows that $\sigma_b\sigma_c$ gives the area of the circle radius 1, a gross overestimate, and also that using $\delta b\delta c$ instead gives a gross underestimate of the area of the ellipse.

The solution is to take linear combinations of the parameters $p_i$ corresponding to a rotation in parameter space, such that the new parameters $v_i$ will have zero correlation or off-diagonal covariance – that is, to change the parameter basis set. These new parameters can be found by diagonalising the parameter covariance matrix. (If the fitting routine used does not return the parameter covariance matrix, it is not difficult to calculate; see Supplementary Information §3). The resulting eigenvectors specify the $v_i$. Repeating the LS fit with these parameters, we get the identical fit and we can recover the original $b_f$ and $c_f$ by a back transform from the fitted values of $v_i$. Plotting $L(v_1, v_2)$ in Fig.1e (contour plot in Fig.1f) we see that now the product of the uncertainties $\sigma_{v_1}\sigma_{v_2}$ is the same as the product $\delta v_1 \delta v_2$ of the widths of the Gaussians $L(v_{1f}, v_2)$ and $L(v_1, v_{2f})$ and either gives us exactly what we require.

To complete the calculation of the BFI, we need the parameter pdfs in parameter space. It is usual to estimate these as top-hat or rectangle functions, of width $\Delta p_i$ and area unity. This is justified as a consequence of Laplace's Principle of Indifference.[9] Then the BFI is the integral $L$ that we have just found, divided by the product of the parameter ranges, $\Delta v_1 \Delta v_2$. Better, though, we do not need to estimate the ranges $\Delta v_i$ if we have already estimated the ranges $\Delta b$ and $\Delta c$, because the product of the ranges is an area in either parameter basis, and areas are invariant under rotation. For discussion of choosing the ranges, see the examples below, and the Supplementary Information §4. Best of all, we do not need to make this transformation of the basis set nor carry out the fit in this basis. The diagonal elements of the parameter covariance matrix in this basis, $\text{Cov}_v$, are the squares of the uncertainties on the parameters, so what we require for the calculation of the BFI in this basis is the product of these elements. Since the off-diagonal elements are zero, this product is the determinant of $\text{Cov}_v$. But the determinant of a matrix is invariant to rotations, and so it is sufficient to take the determinant of the parameter covariance matrix in the $p$ basis, $\text{Cov}_p$. That leads to the principal result reported here, that the BFI of a LS fit with $n$ parameters is given by

$$\text{BFI} = (2\pi)^{n/2} L_{max} \frac{\sqrt{\det \mathbf{Cov_p}}}{\prod_{i=1}^{n} \Delta p_i} \qquad (1)$$

The prefactor in Eq.1 comes from the standard integral of an *n*-dimensional Gaussian. Then the BF between two hypothesis or models is the ratio of their BFIs; more conveniently the logarithms are used and lnBF is the difference of their lnBFIs.

The area defined by the ranges generalises to a volume in parameter space, known as the prior parameter volume. Similarly, the square-root of the determinant of the covariance matrix defines another, smaller volume in the same space, the posterior parameter volume. The ratio of these two volumes has been described as quantifying Occam's Razor, and has been termed the Occam Factor.[9,11]

It is worth noting that Eq.1 is exact, if and only if the functions $L(v_i)$ are Gaussians with widths much less than $\Delta v_i$ as in Fig.1b. It is not difficult to check whether these conditions are satisfied, nor to make reasonable corrections to the BFI (if needed) when they are not. The computational power required for Eq.1 and for checking the $L(v_i)$ or $L(p_i)$ is no more than is required for the LS fit itself. For example, the LS fits with 30 or more parameters in Example 3 below and the calculations of the functions $L(v_i)$ for these fits were originally carried out in a few minutes on a ten-year-old Samsung notebook running Mathematica© 7 under Windows XP.

The other important observation to make is that these methods are applicable to Maximum Likelihood (ML) fitting. In contrast to LS fitting, ML fitting can easily handle the simultaneous fitting of multiple data sets with different error bars, and also it can handle outliers in a rigorous and respectable way. See Example 1 below for both these issues.

**Example 1: Pressure dependence of the band-gap of GaAs.**

The main purpose of this example is to show how the Bayes Factor can be used to decide between two fits which have equal goodness of fit (ln*L* and BIC) to the data. It also shows how ML fitting can be used together with the BF when there are outliers, to obtain much improved parameter values and uncertainties and a new conclusion.

The band-gap of GaAs increases with pressure $P$, as $E_g(P)$, with an initial pressure coefficient that is expected to be about 10 meV kbar$^{-1}$,[12] and with some slight sub-linearity (Fig.1a, which appears to show an increase and then a decrease, but this is because we have subtracted 9.5 meV/kbar to make the curvature more obvious). Six published values from different groups ranged from about 10 to 12 meV kbar$^{-1}$, with curvatures from 0 to –60 μeV$^2$ kbar$^{-2}$.[13] These discrepancies were attributed to the inherent inaccuracies of high-pressure experimentation. However, by comparing all six, we showed that they could be reconciled, with a single value of the pressure coefficient of 11.6 ± 0.2 meV kbar$^{-1}$, by fitting not with the quadratic, $a + bP + cP^2$ but by making the physically-reasonable assumption that the bandgap varies linearly with the density.[13] The density was calculated using the Murnaghan equation of state (see Supplementary Information §5). The key point is that the curvature $c$ then varies with the pressure, decreasing as the bulk modulus $B_0$ increases with pressure as $B_0 + B'P$. The parameter $b$, the initial pressure coefficient of the bandgap, becomes $\Xi/B_0$. In the quadratic fits, $c$ then varies with the pressure range of the experiment, and the covariance between $b$ and $c$ gives the variations in $b$ reported. However, this problem could not be identified or resolved from a single dataset. The Bayes Factor is needed to choose between the two rival models, the polynomial fit and the Murnaghan fit, and can do so from a single dataset.

Fig.1a shows two key data sets. That of Goñi *et al.*[10] has the highest pressure range, yet by inspection, the data clearly cannot give a preference to either fit to it. This is quantified by the BIC. For comparison with ln$L$ and lnBFI values we prefer to use the SBIC = –½BIC (also known as the Schwarz criterion) with values 255.318 for the Murnaghan fit and 255.335 for the polynomial. The usual interpretation of differences of ln$L$ and SBIC, and of the value of lnBF, between two models, is that 1 is barely worth a mention, 2 indicates some evidence for the model with the higher value, 5, strong evidence, and 10, decisive evidence.[3,8] These values correspond to odds of $e$, $e^2$, $e^5$ and $e^{10}$ to 1 on the preferred model, or against the other model.

A $\Delta$SBIC of 0.017 is utterly insignificant. Yet $b = 10.9 \pm 0.17$ meV/kbar, while $\Xi/B_0 = 11.6 \pm 0.27$ meV/kbar. Only the comparison with the other data, particularly that of Perlin *et al.*[14] could previously establish that the latter value was to be preferred. The conclusion from the BF for the Goñi data is quite different. The ranges we pick for $a$ and $b$ or $a$ and $\Xi/B_0$ are not important, for they are the same physical quantities so the ranges are the same for both fits. But $c$ has to be given a range at least as large as the reports in the literature, 60 μeV$^2$ kbar$^{-2}$, which is large compared with the uncertainty on its fitted value of $-13.7 \pm 1.0$ μeV$^2$ kbar$^{-2}$. We use $\Delta c = 100$ μeV$^2$ kbar$^{-2}$. In contrast, $B'$ is known to be about 4 to 5 for very many materials, so we can take $\Delta B' = 2$, which is not so large compared with the error on the fitted value, $4.47 \pm 0.33$. These difference in ranges and uncertainties gives lnBFI = $-254.5$ for the Murnaghan fit, and $-257.5$ for the polynomial – a difference, lnBF, of 3. So we have odds of 20 to 1 on for the Murnaghan fit. That is essentially the same result as in Frogley *et al.*[13], but from a single data-set rather than six, and quantitatively.

We consider briefly the data from Perlin *et al.*[15], to demonstrate the use of the BF in ML fitting. This data had the smallest pressure range of the published data-sets, but was very precise. An LS fit gives $a$ and $b$ or $\Xi/B_0$ quite accurately, while $c$ and $B'$ have large uncertainties as the pressure range is insufficient to get the curvature accurately. The best approach is to fit this data together with the Goñi *et al.*[10] data. This is straightforward using ML fitting. We write the combined ln$L$ as the sum of the probabilities of the residuals, while attributing different pdfs and perhaps different parameters to the Perlin residuals and the Goñi residuals (see Supplementary Information §5). Moreover, by inspection we can see that the Perlin data has outliers, which can distort LS fits. In ML, these are dealt with by a pdf with fatter tails than the Gaussian pdf.[2] The ln$L$ function is then maximised with respect to the parameters using the Mathematica© `FindMaximum` function. This Mathematica function does not return the parameter covariance matrix, nor the uncertainties on the best-fit parameter values, so we calculate them by constructing the Hessian matrix of the second differentials of $L$ with respect to the parameters. See Supplementary Information §3. The desired covariance matrix is minus the inverse of the Hessian, and the uncertainties on the fitted values are the square-roots of the diagonal elements of the covariance matrix. In this way, the lnBFI of the joint fit with Gaussian pdfs and all six fitting parameters is –435.0. However, using a double-Gaussian pdf for the Perlin data the lnBFI improves to –425, i.e. lnBF = 10 – a decisive preference. Using the literature value of $B'$ for both datasets,[15] instead of fitting it, makes little difference to the lnBF but improves the uncertainties on $E_0$ and $\Xi$. Finally, using the same $\Xi$ for both worsens the lnBFI down to –430, lnBF = –6 relative to the preferred fit. This is a strong indication, with odds of $e^6 = 400$ to one, that the Perlin and Goñi datasets have different pressure coefficients, of 11.28 ± 0.04 and 11.6 ± 0.06 meV/kbar respectively.

This is a conclusion – and a precision – that Frogley *et al.*[13] could not reach. It is physically reasonable, as Goñi *et al*. used the absorption edge to identify the band-gap, while Perlin *et al*. used luminescence, which is expected to go slightly deeper below the band-gap with pressure as the electron effective mass increases, thus displaying a slightly lower pressure coefficient than the band-gap.

**Example 2: Muon anti-level-crossing signals.**

This example illustrates how the BF encourages the inclusion of more physically-meaningful parameters in a model than do the BIC or the traditional 3σ or other classical tests. Fig.2a shows an anti-level crossing (ALC) spectrum observed in muon spin spectroscopy from an organic molecule.[16] Theory permits optical excitation to affect the peak position, the width and the strength of an ALC.

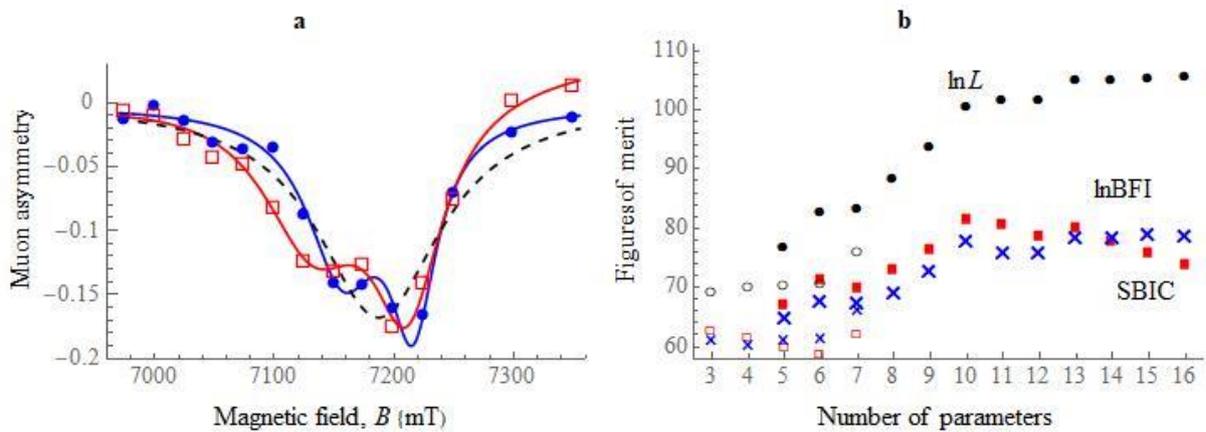

**Fig.2. Muon spin spectroscopy**. Data from an experiment, for muon polarisation as a function of magnetic field, is shown in (a).[16] Error bars on the data are estimated at ± 0.015. A linear background function has been subtracted from the data. The solid blue datapoints (•) were recorded in the dark, while the open red datapoints (□) were photo-excited. The broken black line shows a fit of a single Lorentzian peak to all the data. The blue and red solid lines show fits of two Lorentzian peaks with all parameters fitted separately in the dark and in the light. In (b), the evolution of the figures of merit of the fit with the number $p$ of fitting parameters is shown. SBIC = –½BIC is plotted to fit on the same ordinate. The open or small data points from three to seven parameters are for a single peak while the solid or large datapoints from five to 16 parameters are for two peaks.

In the field region over which the ALC measurements are carried out, the muon's spin may not be fully polarised and there will be a background from the positron detection.[17] Wang *et al.*[16] did not accurately quantify whether two peaks are resolved in the photoexcitation cell, although two overlapping peaks are resolved when using a standard Ti cell. They also did not determine (from LS fitting) whether any effect of photoexcitation is on the peak position, area or width of one or both peaks, although they did undertake a model independent integration of the data, which demonstrated a change in amplitude. ALC peaks are expected to be Lorentzian, there may be one or two peaks in their data set and a linear background function, so anything from 3 to 16 parameters may be needed. Fig.2a shows the three-parameter fit of a single peak with no photosensitivity which returns SBIC = 63, lnBFI = 61 (for details, see Supplementary Information §6). The full sixteen-parameter fit for two peaks (six parameters, plus six more for photosensitivity) with linear baselines (four more parameters) returns SBIC = 74, lnBFI = 82, so lnBF = 21, and is therefore highly preferable on both counts, BIC and BF.

It is interesting that fitting with a single peak shows no light-induced effect. As parameters for the light-induced changes are included for the peak position, width and area of the single peak (open and small data points for $p = 4 - 6$ in Fig.2b), the SBIC (–½BIC) decreases and the lnBFI figure of merit scarcely increases. It is only with the inclusion of one background term ($p = 7$) that either figure of merit shows any substantial increase. In contrast, two peaks (solid or large data points for $p = 5 - 16$ in Fig.2b) shows substantial increases in the quality of fit (ln$L$) and in the SBIC and lnBFI when parameters for light-changes are included in the fit. The SBIC peaks at 81 at 10 parameters and then decreases, discouraging the remaining six. The lnBFI is at nearly its maximum at $p = 10$, but does reach its maximum of 82 for $p = 15$ and $p = 16$.

Only seven parameters pass the 3σ test. The other nine are all within 1½σ of zero and would conventionally be excluded from the model (i.e. set to zero). The BF gives a different message. It encourages the inclusion of all the physically-meaningful parameters. This is partly because their ranges are small, and partly because their inclusion increases the errors on the other parameters when the correlation coefficients are non-zero. Setting parameters to zero gives spurious decreases in the uncertainties of the remaining parameters and spurious shifts of their fitted values in exactly the same way as fixing any fitting parameter at a non-zero round number within error of its fitted value would do. In the case of the muon ALCs, theory permits light to shift any of the peak parameters, if only by little, and so the most reliable estimates of all the parameters are those given by the fit that includes them all.

As an example, the increase in width of the 7150 gauss peak under photoexcitation is given as $21 \pm 6$ gauss at $p = 10$ but as $16 \pm 13$ gauss at $p = 16$. The latter estimate is less precise but more reliable – better – a more accurate description of what the data says.

**Example 3: Carbon nanotube Raman spectrum**

This example demonstrates how the Bayes Factor provides a quantitative answer to the problem, whether we should accept a lower quality of fit to the data if the parameter set is intuitively better. It also provides a simple example of a case where the BF should be termed the RBF and given some corrective treatment.

The dataset is a Raman spectrum of the radial breathing modes of a sample of carbon nanotubes under pressure, previously fitted in the traditional way.[18] On the way to a definitive fit, we monitor the quality of fit while parameters are added in four models (for details, see Supplementary Information §7). All models have seven sharp pseudo-Voigt peaks that are clearly seen in the data (Fig3a; the two strong peaks are clearly doublets), a constant for dark current, and a factor describing the Gaussian content in the pseudo-Voigt peak shape, amounting to 23 parameters. In the Fourier model we add a Fourier background to improve the fit. Equivalently, in the LagP model, we use a polynomial background using the Lagrange polynomials $P_i(x)$. In the Peaks model we add extra broad peaks as background. All three models gave good fits; however, it was noticeable that the three backgrounds all looked as if they were related to the sharp peaks, rather like a heavily broadened replica (see Supplementary Information, Fig.S2).

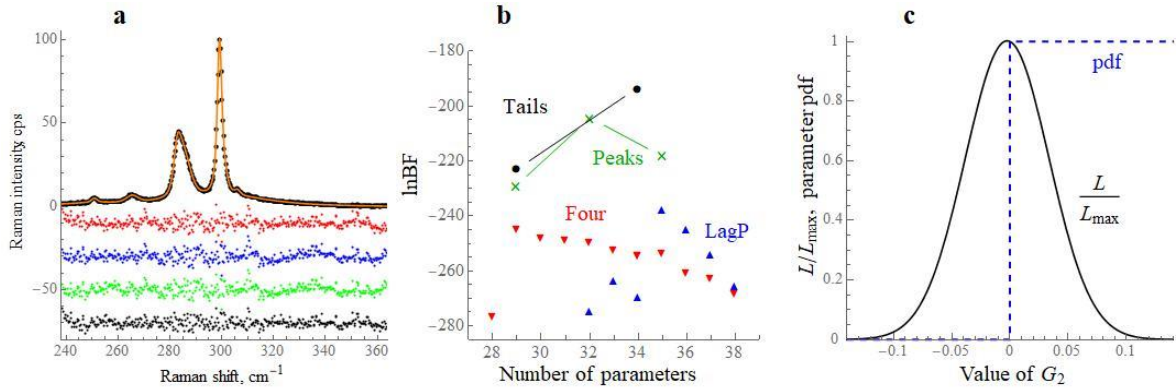

**Fig.3a.** The carbon nanotube Raman spectrum (black datapoints). The fit shown (orange line) is the fit using the Tails model. The residuals for four fits are shown, ×10 and displaced successively downwards. The top one is a 29-parameter Fourier fit (red) Then a Peaks 29-parameter fit (blue), a Tails 29-term fit, green), and (bottom) a 37-term Fourier fit (black). In (b), the evolution of the BFIs is shown against the number of parameters; LagP indicates the Lagrange polynomial model. In (c) the normalised likelihood function $L(G_2)$ (solid line) and the pdf of this parameter (dashed line) are plotted.

Accordingly, in a fourth model, we modify the peak shape, giving it stronger, fatter tails than the pseudo-Voigt peaks (Tails model). The stage shown in Fig.3 is chosen for discussion because, fortuitously, we have an equal number, 29, of fitting parameters in the Fourier, Peaks and Tails models, and quite good fits with all three, with ln$L$ values –65, –67, –70 respectively. (For a ln$L$ of –66 the LagP model needs 33 parameters.) With the same number of parameters, the SBIC values vary as the ln$L$ values (–78, –80, –83 respectively).

(The polynomial LagP model is significantly disfavoured at –85). So the Fourier fit appears to be the best.

The Bayes factors tell a very different story. The Tails model has the best value, at lnBFI = –223. The Peaks fit comes close, at lnBFI = –229 or lnBF = –6. The Fourier background fit is decisively rejected, with lnBFI = –244 or lnBF = –21. A further important consideration is that the Tails model has a limited number of parameters that could usefully be added, and does not span the data space. There are many possible datasets which it could not fit. Intuition suggests that because it does fit this data set, it is probably the true model. In contrast, the number of extra parameters in the other three is unconstrained and by adding sufficient, up to $d$, we are certain to achieve a perfect fit to the data. The actual number used is therefore itself a fitting parameter, with an uncertainty perhaps of the order of ±1 or less, and a range from 0 to perhaps a quarter or a half of the number of data points $d$. We may therefore add to lnBF the factor $\sim \ln 4d^{-1} \sim -5$ for a few hundred data points. This takes Peaks to $\Delta \ln BF = -11$, which is strong evidence to reject this model too. This quantifies the intuition that a model not guaranteed to fit the data, but which does, is preferable to a model which certainly can fit the data because it spans the data space.

Taking all models further, more Fourier or polynomial terms, more extra peaks, and more parameters distinguishing the tails in Tails, Tails reaches only ln$L$ = –35. The other models achieve ln$L$ values close to –20. However, Fig.3b shows that this reduction by the factor of $e^{15}$ in the probability of the data under the fit is more than compensated by the greater increase of the probability of the parameters. Adding a further 5 for not spanning the data space, Tails has a decisive BF of +16 against the Peaks model and +50 or +55 against the Lagrange polynomial and Fourier models.

Finally, Fig.3c shows a case earlier in the fitting, where we gave each of the seven peaks its own fraction $G_i$ of Gaussian content in the pseudo-Voigt peak shape. Fitted values were sometimes outside the meaningful range of $0 \leq G_i \leq 1$, so this condition was added as a constraint. Fig.3c shows that in this case, with a fitted value of $G_2 = 0$, half of the Gaussian function $L(G_2)$ is outside the range of non-zero probability for this parameter. Its contribution to the BFI integral is thus twice what it should be. This can be corrected by dividing the BFI by 2, or subtracting ln2 from lnBFI and lnBF. Such corrections are not always exact, and then it would be appropriate to call the outcome the RBF, not the exact BF.

**Discussion:**

The examples are drawn from solid-state physics because that is our discipline so we have ready access to data from it. Data from any other fields that use LS or ML fitting could be handled as readily, e.g. Rietveld refinement of X-ray diffraction data, linear and nonlinear regression in econometrics or psychology – wherever there is an analytic model to fit a data set. Rate-equation models such as SIR and its refinements in epidemiology sometimes have analytic solutions and then their fits to data can be handled. Generally, though, rate-equations model do not have analytic solutions and it is not clear if there is an easy way to calculate the BF in those cases. To conclude, the central point of this paper is to show that the BFI and the BF can readily be calculated after all and any LS and ML fits, and, by quantifying our intuitive discrimination among parameters, give valuable information for model selection that traditional figures of merit do not. For parameter estimation, using the preferable model may give estimates which are less precise, but which are certainly more accurate.

**Acknowledgements:** We thank Brian Smith, Ian Bell and colleagues at Renishaw plc for discussions. Graham Thompson, Adrian Bevan and Jonathan Hays of the Particle Physics

Research Group at Queen Mary assisted with ML methods. A.J.D. acknowledges funding from the European Research Council, from the project MuSES.

**Author Contributions:** D.J.D. initiated this study and completed it. J.C. found the Bayes Factor literature and implemented many of the calculations. A.J.D. provided the muon example, and critically discussed all results presented. All authors contributed to the final manuscript.

**Competing Interests:** The authors declare no competing interests.

**References**

1. Fuller, W.A. *Measurement Error Models*. (Wiley-Blackwell, Oxford, 1987).

2. Sivia, D.S. & Skilling, J. *Data Analysis*: *A Bayesian Tutorial*. (Oxford University Press, Oxford, 2006).

3. Jeffreys, H. *Theory of Probability* (Oxford University Press, 1939, 1948, 1961, 1979).

4. Jaynes, E.T. Bayesian methods: General background. An introductory tutorial in *Maximum Entropy and Bayesian Methods in Applied Statistics* (ed. Justice, J.H.) pp. 1-25 (Cambridge University Press, 1985).

5. Faulkenberry, T.J. Computing Bayes factors to measure evidence from experiments: An extension of the BIC approximation. *Biometrical Lett.* **55**, 31-43 (2018).

6. Wagenmakers, E.-J. A practical solution to the pervasive problem of *p* values. *Psychonomic Bull. Rev*. **14**, 779-804 (2007).

7. Jarosz, A.F. & Wiley, J. What are the odds? A practical guide to computing and reporting Bayes Factors. *J. Problem Solving* **7**, Art.2 (2014).

8. Kass, R.E. and Raftery, A.E. Bayes Factors. J. Am. Stat. Ass. **90**, 773-795 (1995)

9. Gull, S.F. Bayesian inductive inference and maximum entropy in *Maximum Entropy and Bayesian Methods in Science and Engineering* (Kluwer Academic Publishers, 1988), ed. Erickson G.J. and Smith C.R., vol.1, pp 53-74.

10. Goñi, A.R., Strössner, R.K., Syassen, K. & Cardona, M. Pressure dependence of direct and indirect optical absorption in GaAs. *Phys. Rev.* B**36**, 1581-1587 (1987).

11. Rasmussen, C.E., Ghahramani, Z. Occam's Razor in *Advances in Neural Information Processing Systems* 13 (MIT Press, Cambridge MA, 2001), ed. Leen, T.K., Dietterich, T.G. and Tresp, V.

12. Paul, W. & Warschauer, D.M. *Solids under Pressure* (McGraw-Hill, New York, 1963).

13. Frogley, M.D., Sly, J.L. & Dunstan, D.J. Pressure dependence of the direct band-gap in tetrahedral semiconductors, *Phys. Rev.* B **58**, 12579-12582 (1998).

14. Perlin, P., Trzeciakowski, W., Litwin-Staszewska, E., Muszalski, J. and Micovic, M. The effect of pressure on the luminescence from GaAs/AlGaAs quantum wells. *Semicond Sci Technol.* **9**, 2239-2246 (1994).

15. McSkimin, H.J., Jayaraman, A. and Andreatch, P. Elastic moduli of GaAs at moderate pressures and the evaluation of compression to 250 kbar. J. Appl. Phys, **38**, 2362 (1967).

16. Wang, K., Murahari, P., Yokoyama, K., Lord, J.S., Pratt, F.L., He, J., Schulz, L., Willis, M., Anthony, J.E., Morley, N.A., Nuccio, L., Misquitta, A., Dunstan, D.J., Shimomura, K., Watanabe, I., Zhang, S., Heathcote, P. & Drew, A.J. Temporal mapping of photochemical reactions and molecular excited states with carbon specificity. *Nature Materials* **16**, 467-473 (2017).

17. Yokohama, K., Lord, J.S., Murahari, P., Wang, K., Dunstan, D.J., Waller, S. P., McPhail, D. J., Hillier, A.D., Henson, J., Harper, M.R., Heathcote, P. & Drew, A.J.

The new high field photoexcitation muon spectrometer at the ISIS pulsed neutron and muon source. *Rev. Sci. Instrum.* **87**, 125111 (2016).

18. Torres-Dias, A.C., Cerquiera, T.F.T., Cui, W., Marques, M.A.L., Botti, S., Machon, D., Hartmann, M.A., Sun, Y., Dunstan, D.J. & San-Miguel, A. From mesoscale to nanoscale mechanics in single-wall carbon nanotubes. *Carbon* **123**, 145-150 (2017).

# Easy computation of the Bayes Factor to fully quantify Occam's razor

D.J. Dunstan,* J. Crowne, and A.J. Drew

School of Physics and Astronomy, Queen Mary University of London, London E1 4NS, UK.

## Supplementary Information

**§1. Parameter Covariance Matrix:** Some least-squares fitting routines return the parameter covariance matrix **Cov**. When that is not so, it can be readily calculated. The residuals are expressed as functions of the parameters $p_i$, and then the likelihood $L$ can be calculated (§2). Then the fitted values for all parameters except one or two can be substituted in to give the $m$ functions $L(p_i)$ and the $\frac{1}{2}(m^2 - m)$ functions $L(p_i, p_j)$. From these, the second derivatives can be calculated quite quickly. The Hessian matrix is

$$\mathbf{H} = \begin{pmatrix} \frac{\partial^2 L}{\partial p_1^2} & \cdots & \frac{\partial^2 L}{\partial p_1 \partial p_m} \\ \cdots & \cdots & \cdots \\ \frac{\partial^2 L}{\partial p_1 \partial p_m} & \cdots & \frac{\partial^2 L}{\partial p_m^2} \end{pmatrix} \quad (S1)$$

and then $\mathbf{Cov} = -\mathbf{H}^{-1}$.

**§2. Least-squares and Likelihood:** Let the probability (pdf) that a residual is of magnitude $r$ be

$$P(r) = \frac{1}{\sqrt{2\pi\sigma^2}} e^{-\frac{r^2}{2\sigma^2}}$$
$$L = \prod_{i=1}^{n} P(r_i) \quad (S2)$$
$$\ln L = -n \ln \frac{1}{\sqrt{2\pi\sigma^2}} - \sum_{i=1}^{n} \frac{r_i^2}{2\sigma^2}$$

so that minimising the sum of the squares of the residuals is the same thing as maximising $L$ or $\ln L$.

**§3. Multiple integral for Bayes factors:** The Bayes factor integral BFI for a model with $m$ parameters $p_i$ with flat pdfs across the parameter ranges $\Delta p_i = p_{ihi} - p_{ilo}$, is given by

$$\mathrm{BFI} = \frac{1}{V_R} \int_{p_{1lo}}^{p_{1hi}} \int_{p_{2lo}}^{p_{2hi}} \cdots \int_{p_{plo}}^{p_{phi}} L(p_1, p_2 \ldots p_m) dp_1 dp_2 \ldots dp_m \quad (S3)$$
$$\text{where} \quad V_R = \prod_{i=1}^{m} \Delta p_i$$

**§4. Ranges of parameters:** Ranges should be set *a priori*, using all knowledge that may be available, but before any inspection of the data. No great precision is needed, since it is the logarithms of the ranges that matter. Choosing the ranges has been criticized as introducing subjectivity; however, the key point is that the ranges are given quantitatively and should be justified.[2,9] The values used can be defended or opposed on objective grounds. For physically-meaningless parameters such as polynomial coefficients, it is helpful to consider the data rescaled to the ranges 0 to 1 in $x$ and $y$ (that can be done automatically without inspecting the data). For example, if some data are to be fitted with $a + bx + cx^2$ and there is no prior information about the parameters, when the data are rescaled to the ranges $0 - 1$ on

both axes, all parabolae that could fit in this would be encompassed by the ranges $0 \leq a \leq 1$, $-4 \leq b \leq 4$ and $-4 \leq c \leq 4$. The rescaled data can be fitted and these ranges used, or these ranges can be rescaled for the original data.

For the fits shown in Fig.1 we set $a = 0$ in the polynomial fit $a + bx + cx^2$ so as to have a two-parameter problem for Fig.1. The Murnaghan fit uses Eq. S4, where the pressure coefficients $\Xi/B_0$ and $b$ are expected from theory and experiment[12] to be about 10 meV kbar$^{-1}$ and the pressure derivative of the bulk modulus $B'$ is expected from experiment in many materials and from theory to be about 4 to 5. So their ranges were chosen as 2 meV kbar$^{-1}$ and 2 respectively, while $c$ as mentioned in the main text was given a range of 60 μeV$^2$ kbar$^{-2}$. In Example 1, where we fit for the $y$-axis intercept $a$ as well, we give this a range of 100 meV.

For the muon data, Fig.2, we fit first with a single Lorentzian peak with the three parameters $P_1$, $W_1$ and $A_1$. $W$ is the half-width at half-height. The area $A$ of a peak of height 1 and $W = 1$ is $\pi$. Then with data rescaled to $0 - 1$ on both axes and assuming most of the peak (to say $2W$ each side of the peak) is included in the dataset, $\Delta P_1 = 1$, $\Delta W_1 = ¼$, $\Delta A_1 = ¼\pi$ are first estimates. This would correspond to unscaled ranges of 400, 100 and 80. All three could be argued down, $\Delta P_1$ on the grounds that the data would not have been selected to have the peak far from the middle, $\Delta W_1$ and $\Delta A_1$ on the grounds that the data surely extends over several peak-widths. The light-induced changes in the parameters $\delta_{Lp}$ may be given similar ranges, or their ranges may be argued down on the grounds that large changes would have been identified without recourse to these refined fitting procedures, or on the grounds that large changes are not expected. We settled on the values, $\Delta P_1 = 400$ but $\Delta P_2 = 100$, $\Delta W_i = 50$, $\Delta A_i = 40$. Then for the light-induced effects, assumed to be small, we chose $\Delta\delta_{LP} = 100$, $\Delta\delta_{LW} = 50$, $\Delta\delta_{LA} = 40$. Finally, for the residual background parameters (see below), we chose 0.03.

For the Raman data, Fig3, the ranges for the 23 parameters (seven peak, seven widths, seven areas, plus a pseudo-Voigt Gaussian proportion $G$ and a zero offset to take account of dark current) common to the three models do not affect the comparison of the models, so it is not important what values are chosen. We used 120 cm$^{-1}$ for the Raman peak positions as they must fall in the range of the data, 238 – 364 cm$^{-1}$. For their widths, 10 cm$^{-1}$, on physical grounds (we know they are nanotube RBM modes). For the areas, 2500 cm$^{-1}$, because this is the area of a peak with the maximum width with the greatest Raman intensity of 100 in the data. The zero offset (dark count rate) is expected to be of the order of 1. The fraction of Gaussian in the Voigt lineshape is necessarily in the range 0 to 1. Much more important are the ranges of the parameters not common to the two models. With no information to justify physically the properties of extra background peaks, they could have any position, so their range is again 120 cm$^{-1}$, their widths could be up to 100 cm$^{-1}$ and areas up to 25000 cm$^{-1}$. The Fourier and polynomial background terms have no physical interpretation, and their coefficients may have values comparable with the data values, so a range of 100.

### §5. Example 1: Pressure dependence of the GaAs bandgap.

The Murnaghan equation of state derives from standard interatomic potentials and finite elasticity theory, using the approximation that the pressure derivative $B'$ of the bulk modulus $B_0$ is constant. Then the equation for the volume as a function of pressure is

$$\frac{\Delta V}{V_0} = \left(1 + \frac{B'P}{B_0}\right)^{-\frac{1}{B'}} - 1 \tag{S4}$$

Assuming that the bandgap of GaAs, $E_g$ is linear with the density,[13]

$$\Delta E_g = \Xi \left(1 + \frac{B'P}{B_0}\right)^{\frac{1}{B'}} - 1 \qquad (S5)$$

Fitting the datasets independently with three parameters each ($E_0$, $\Xi/B_0$ and $B'$ with ranges as discussed in §4) the Goñi data gives $\Xi/B_0 = 11.58 \pm 0.3$ and $B' = 4.5 \pm 0.3$. the lnBFI is –254.4. The Perlin data gives $\Xi/B_0 = 11.55 \pm 0.2$ but $B' = 6.6 \pm 2.5$, outside the range we adopted. This makes the BFI incorrect. The problem is that the Perlin data covers too small a pressure range to determine $B'$ accurately. The remedy is to give $B'$ a much larger range, to express our ignorance, not of constraints on real value of $B'$, but of what data with noise may return. We take $\Delta B' = 10$, and that gives lnBFI = –180.7. The total lnBFI for the two fits together is the sum, –435.0.

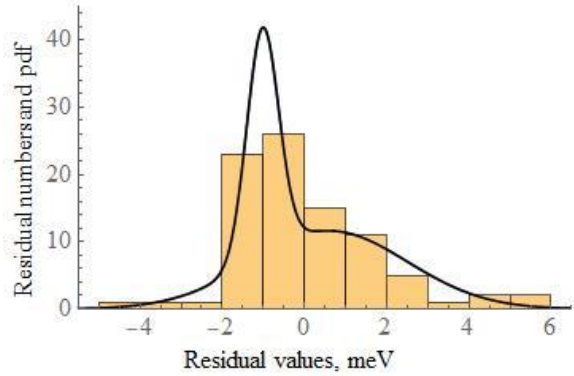

Fig.S1. A histogram of the residuals returned by the least-square fit to the Perlin *et al.*[14] data. The double Gaussian pdf used in the maximum likelihood fitting of the data is also plotted, multiplied by the number of data points (88).

In both the data (Fig.1a) and in the residuals of the Perlin fit (Fig.S1), it is clear that there are outliers, particularly to the right. In ML fitting of the Perlin data alone, we use the first product term of Eq.S6 below, in which the outliers are accommodated by a broader Gaussian contributing a fraction $f$ of the residuals pdf (Fig.S1). This gives a very substantial improvement in the quality of fit to data, with the ln$L$ improved from –171 to –159. So, for the ML fitting of the Perlin data[14] (datapoints $i = 1$-88) together with the Goñi data (datapoints $i = 89$-146),[10] the likelihood is written as

$$L = \prod_{i=1}^{88} \left( \frac{1-f}{\sqrt{2\pi\sigma_P^2}} \exp\left(-\frac{r_i^2}{2\sigma_P^2}\right) + \frac{f}{\sqrt{2\pi\sigma_O^2}} \exp\left(-\frac{r_i^2}{2\sigma_O^2}\right) \right) \prod_{i=89}^{146} \frac{1}{\sqrt{2\pi\sigma_G^2}} \exp\left(-\frac{r_i^2}{2\sigma_G^2}\right) \qquad (S6)$$

where as well as the fitting parameters in the residuals, the proportion of outliers, $f$, and the standard deviations of the outliers, $\sigma_O$, of the rest of the Perlin data, $\sigma_P$, and of the Goñi data, $\sigma_G$, are included as fitting parameters. Moreover, since the Goñi data give a $B'$ in excellent agreement with the literature value, we use that instead of fitting it. Then fitting with a single $\Xi$ for both datasets yields $\Xi/B' = 11.36 \pm 0.03$ meV kbar$^{-1}$ and lnBFI = –429.5. Fitting with separate pressure coefficients for the two data sets gives $\Xi_P/B' = 11.28 \pm 0.04$ meV kbar$^{-1}$, $\Xi_G/B' = 11.59 \pm 0.06$ meV kbar$^{-1}$ and lnBFI = –425.0. So lnBF = 4.5, strongly in favour of the different pressure coefficients for the two datasets.

## §6. Example 2: Muon data fits.

Lorentzian peaks were used, with peak positions $P_i$, widths $W_i$ and areas $A_i$:

$$y(x) = \frac{A}{\pi}\frac{W}{W^2+(x-P)^2} \tag{S7}$$

Light-induced effects were added by adding to each parameter $p$ a term $z\delta_{Lp}$, where $\delta_{Lp}$ is the change in $p$ under illumination, and writing the data in 3D as $y_i(x_i,z_i)$ with $z_i = 0$ for data in the dark and $z_i = 1$ for the photoexcited data. Linear baselines were added, defined by the four corner points, $y_{00}$ at $(x_{lo},0)$, $y_{x0}$ at $(x_{hi},0)$, $y_{0z}$ at $(x_{lo},1)$ and $y_{xz}$ at $(x_{hi},1)$. This enabled least-squares fitting to all data simultaneously, using the Mathematica function `NonlinearModelFit[data, func, {x,z}]`. The first fit used only the three parameters $P_1$, $W_1$ and $A_1$. Adding successively the other thirteen parameters, the highest lnBFI is found when all sixteen are included. To prepare Fig.2b, the other parameters were added in order of their final importance, first $\delta_{LP}$, $\delta_{LW}$, $\delta_{LA}$ and $y_{xz}$ (small and open data points in Fig2). Then for two peaks, $\delta_{LP}$, $\delta_{LW}$ and $\delta_{LA}$ were deleted and $P_2$, $W_2$, $A_2$, $\delta_{LP2}$, $\delta_{LA2}$, $\delta_{LW2}$, $\delta_{LP1}$, $\delta_{LW1}$, $\delta_{LA1}$, $\delta_{Lyx0}$, $\delta_{Ly0x}$, $\delta_{Lyzx}$ successively added. From Fig.2, the SBIC (–½BIC) peaks at 10 parameters, up to and including $\delta_{LW2}$, but discourages the inclusion of any light-induced effect on peak 2 and three of the four corners of the background.

### §7. Nanotube Raman fit.

The data was first fitted using just the seven pseudo-Voigt peaks – 21 parameters, plus a single factor for the proportion of Gaussian in the Voigt lineshape. The fit to this basic model was very poor. It was improved in four ways, as seen in Fig3b. Adding a Fourier $y_F$ or polynomial $y_P$ background (we used the Lagrange functions $P_n$), given by

$$y_F = a_0 + \sum_n \left(a_n \cos\frac{2\pi n(x-x_0)}{(x_{max}-x_{min})} + b_n \cos\frac{2\pi n(x-x_0)}{(x_{max}-x_{min})}\right)$$

$$y_P = a_0 + \sum_n P_n\left(\frac{x-x_0}{(x\_max-x\_min)}\right) \tag{S8}$$

with $x_{min} = 238$ cm$^{-1}$, $x_{max} = 301$ cm$^{-1}$ and $x_{max} = 364$ cm$^{-1}$, gave much better fits. Following the traditional fitting method of adding (broad background) peak also fitted well (Fig.3a). At this point, it was remarked how these fits were giving backgrounds apparently closely related to the sharp Raman peaks (Fig.S1). It looks as if the background is a much-broadened version of the sharp peak spectrum. This suggests a physical origin for the background, that in fact the Raman lineshape should have stronger tails. Accordingly, the lineshape was modified by adding a fraction of a much broader Lorentzian function. This also fitted well Fig (3a).

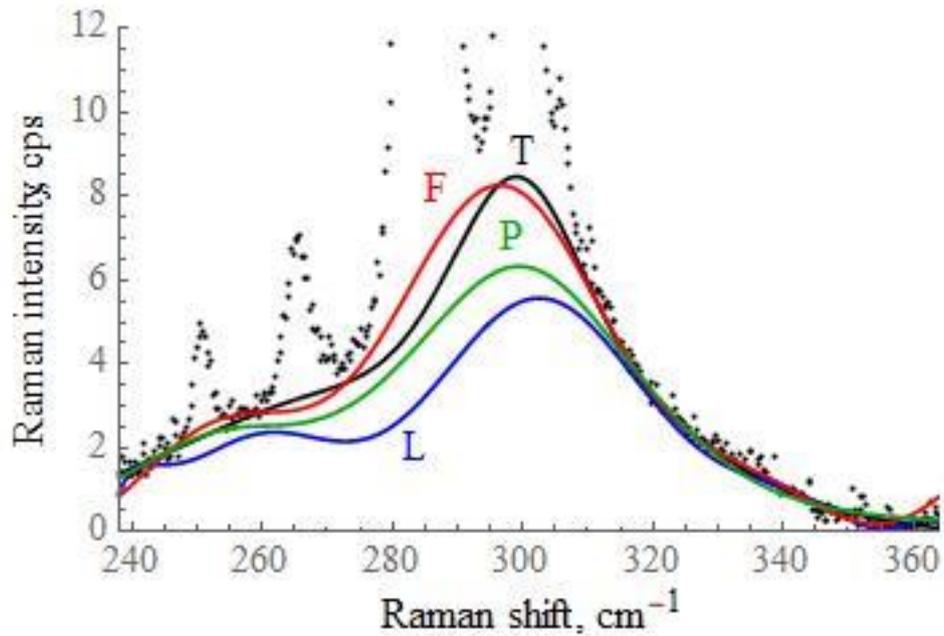

**Fig.S2.** The backgrounds of the four fits discussed are shown wo fits to the carbon nanotube Raman spectrum (black data points). In both fits, seven sharp Voigt peak functions are fitted (dashed blue lines), and the fit residuals are shown, displaced ten units down and ×10. In (a), a background is used, shown ×10 (thin solid lines), consisting of an eighth broad Voigt peak and a three-term Fourier series. In (b), no background is used; instead the seven peak functions are modified to have stronger tails.

Taking all models further, with more terms in the Fourier and polynomial backgrounds, ln$L$ reaches about –20 (best fits to data) and SBIC up to –140 (preferred models). With more background peaks, ln$L$ reaches –20 with a higher SBIC of –130. In contrast, the tails model reaches only ln$L$ = –35 but with a SBIC of –138.